\documentclass[reprint,amsmath,amssymb,aps]{revtex4-1}
\usepackage{graphicx}  
\usepackage{dcolumn}   
\usepackage{bm}        
\usepackage{amssymb}   
\usepackage{floatrow}
\usepackage{mathrsfs}
\usepackage{amsmath}
\usepackage{verbatim}
\begin{document}

\title
{
Piezoelectric Polar Nano Regions and Relaxation-Coupled Resonances\\ in Relaxor Ferroelectrics}

\author{Ling Cai}
\author{Radha Pattnaik}
\author{Joel Lundeen}
\author{Jean Toulouse}
\affiliation{%
Physics Department, Lehigh University, Bethlehem, PA, 18015
}%

\date{\today}

\begin{abstract}
\indent It is a generally accepted fact that the unique dielectric properties of relaxor ferroelectrics are related to the formation of polar nanoregion (PNRs). Less well recognized is the corollary that, because they are polar and therefore lack inversion symmetry, PNRs are also piezoelectric at the nanoscale and can therefore behave as nanoresonators. Using the particular relaxor ferroelectric K$_{\tt1-x}$Li$_{\tt x}$TaO$_{\tt 3}$ (KLT), we show that, when electrically excited into oscillation, these piezoelectric nanoresonators can drive macroscopic electro-mechanical resonances. Unexpectedly however, pairs of coupled resonances are observed, with one of the two exhibiting a characteristic Fano-like lineshape. The complex resonance spectra can be described equally well by two alternative but complementary models both involving two resonances coupled through a relaxation: a purely classical one based on two coupled damped harmonic oscillators and a semi-classical based on two discrete excitations coupled to each other through a continuum. Together, they provide complementary perspectives on the underlying physics of the system. Both reproduce the rapid evolution of the resonance spectrum across three wide temperature ranges, including a phase transition range. In the high temperature range, the coupling between modes is due to the collective $\pi$ relaxation of the lithium ions within PNRs and in the phase transition range to "heterophase relaxation" of the surrounding lattice between its high temperature cubic and low temperature tetragonal phases. The coupling is suppressed in the intermediate range of the collective $\pi/2$ relaxation of the lithium ions. Incidentally, the measured dielectric spectra are shown to bear a surprising but justifiable resemblance to the optical spectra of certain atomic vapors that exhibit electromagnetically induced transparency.

\begin{description}
\preprint{APS/123-QED}
\item[PACS numbers]
 {Piezoelectricity and electromechanical effects}
 {Composites (nanosystems embedded in a larger structure)}
 {Effects of atomic coherence on propagation, absorption, and amplification of light; electromagnetically induced transparency and absorption}
\end{description}
\end{abstract}

\maketitle

\section{\label{sec:level1}Introduction}

K$_{\texttt1-x}$Li$_{\texttt x}$TaO$_{\texttt 3}$\,(KLT), KTa$_{\texttt1-x}$Nb$_{\texttt x}$O$_{\tt 3}$\,(KTN), PbMg$_{\texttt 1/3}$Nb$_{\texttt 2/3}$O$_{3}$-PbTiO$_{3}$\,(PMN-PT), PbZn$_{\texttt 1/3}$Nb$_{\tt 2/3}$O$_{\texttt 3}$-PbTiO$_{3}$\,(PZN-PT), PbSc$_{\texttt 1/2}$Nb$_{\texttt 1/2}$O$_{3}$\,(PSN) belong to the family of relaxor ferroelectrics (RF).  RFs are highly polarizable mixed compounds in which one or several atoms in the unit cell are off-centered already in the paraelectric phase, forming a dipole that can reorient between several crystallographically equivalent directions. At lower temperatures, interactions between off-center ions in the highly polarizable lattice result in the formation of lower symmetry (i.e. permanent) polar nano-regions (PNR)\cite{Jeong}, the size of which can be estimated from neutron and x-ray elastic diffuse scattering\,\cite{Xu, Yong}. Although the term polar nano-regions or PNRs is widely used in the RF literature, it would be preferable in the present paper to label these polar nano-domains or PNDs \cite{PNRs/PNDs} so as to emphasize the long-lived or permanent character of the local distortion and lower local symmetry of these regions below a certain temperature, $T^{*}$, which is essential in explaining the results reported here. Due to their mixed composition and resulting complex structural features, RFs exhibit unique local as well as lattice dynamics, the most characteristic feature of which is the strong frequency dispersion of their dielectric constant commonly identified as ``the relaxor behavior''\,\cite{Toulouse1}. This dispersion is due to the relaxation of the PNRs between different orientations. Simultaneously, when subjected to relatively small dc electric fields, relaxor ferroelectrics\,(RF) exhibit unusual electro-mechanical resonances\,(EM)\,\cite{Pattnaik1,Tu} that are clearly associated with the presence of these PNRs coupling polarization and strain. Similar resonances have also been observed in nanocomposites\,\cite{Sharma} and are interesting for two reasons. First, they provide a sensitive tool to probe the interplay between local and lattice dynamics which is at the core of the behavior of these complex solids. Secondly, they form the basis for the primary applications of RFs in transducers and actuators\,\cite{Park, Noheda}. In an earlier paper\,\cite{Pattnaik1}, we reported the first observation of new resonances in KTN and KLT and interpreted them as evidence for the formation of permanent polar nanodomains in the "paraelectric" relaxor phase. This earlier paper focused on the primary (broad) resonance as a signature of the PNRs, as we did not at the time yet recognize the importance of the secondary (narrow) resonances and the meaning of its characteristic spectral shape. In the present article, we report the results of a much more complete and quantitative study of these resonances in the relaxor KLT over a wide range of  temperatures and frequencies. Most importantly, we now identify pairs of \emph{coupled resonances}, uncover their origin and describe the coupling mechanism that gives rise to their characteristic spectral shapes. These spectral shapes are seen to evolve rapidly with decreasing temperature, first due to their interaction with the relaxations mentioned above, and then to the occurrence of a phase transition. In the vicinity of the phase transition, they provide evidence for the existence of "hetero-phase fluctuations" of the system between its high temperature cubic and low temperature tetragonal phase\,\cite{Yukalov}.\\

\indent In KLT, lithium ions are off-centered from the normal crystallographic site by almost 1{\AA}, thus forming electric dipoles that can reorient among six equivalent cubic directions\,\cite{Yacoby,Pattnaik2}. At lower temperature, interactions between off-center lithium ions result in their displacements becoming correlated, leading to a local transition and the appearance of tetragonal polar nanodomains\,(PND) \,\cite{Yong}. PNRs exhibit two distinct types of local  dynamics. First, they can relax between several crystallographically equivalent orientations via collective 90$^\circ$ and 180$^\circ$ jumps of the Li dipoles\,\cite{Knauss,Doussineau1}. Secondly, being polar and therefore lacking inversion symmetry, they are also piezoelectric and can exert a stress on the surrounding lattice to drive a crystal bar into electro-mechanical resonance. Similar resonances have also been observed KTN and PZN. Given the common characteristics of relaxor ferroelectrics (off-center mixed ions and a high polarizability), the results reported below should be indicative of the behavior of other RFs as well, and to provide a more complete picture of the inter-relationship between mesoscopic and macroscopic dynamics in these compounds. With regards to applications, these results may also contribute to a better understanding of the piezoelectric properties of nanocomposites.

\indent In the next Experimental and Results section, we present the dielectric constant and electro-mechanical results on the two crystals studied. In the Analysis section, we fit the resonance spectra with two different theoretical models, providing complementary perspectives on the underlying physics. Finally, in the Physical Model and Discussion section, we describe schematically the underlying physics of the two models and highlight their special meaning.

\section{\label{sec:level2}Experimental Details and Results}

Two KLT single crystals were grown from solution by the slow cooling method at Oak Ridge National Laboratory. The nominal lithium concentrations of these two crystals were x=3.5\% and x=10\%. However, using a formula proposed earlier to calculate concentrations of lithium based on the transition temperature \cite{van der Klink}, we estimate that actual concentrations in our crystals must have been respectively 2.6\% and 4.7\%, both therefore exceeding the critical concentration of $\leq{2\%}$. The crystals were cut along (100) faces in the form of bars with dimensions $8.6\times5.5\times5\,\mathrm{mm}$ and $5.82\times3.33\times2.87\,\mathrm{mm}$ respectively. Metallic electrodes were evaporated on the two largest parallel surfaces of the samples. In order to rule out possible electrode-sample interface effects, different coating/interface conditions, such as sputtered gold, vapor deposited aluminum and painted silver, were tested to ensure that the same dielectric results were obtained. Different grades of surface polish were also tested, from rough to optical grade, and the same dielectric results were obtained in  all cases. Ultimately, aluminum electrodes were used. The samples were held stress-free inside an open cycle cryostat. For the dielectric relaxation measurements, a small ac electric field\,(0.5\,$\mathrm{V/cm}$) was applied across the short dimension\,(thickness) of the crystal sample. The parallel plate capacitance and the loss tangent were measured with a HP4194A network analyzer, sweeping the frequency from 100\,$\mathrm{Hz}$ to 10\,$\mathrm{MHz}$. The measured capacitance was converted to a dielectric constant through the relation $\epsilon'=Cd/A\epsilon_0$, where $C$ is the capacitance, $d$ the sample thickness, $A$ the area of the electrode and $\epsilon_0$ the free space permittivity. The samples were cooled with liquid helium from room temperature to $\sim20\,\mathrm{K}$. The cooling rate was controlled to be on average 0.2\,$\mathrm{K/min}$ but the temperature was equilibrated at each measuring temperature, allowing sufficient time for the sample to reach thermal equilibrium before each measurement, as monitored by the stability of the capacitance value at that temperature.\\

\begin{figure}
\includegraphics[width=7.4cm]{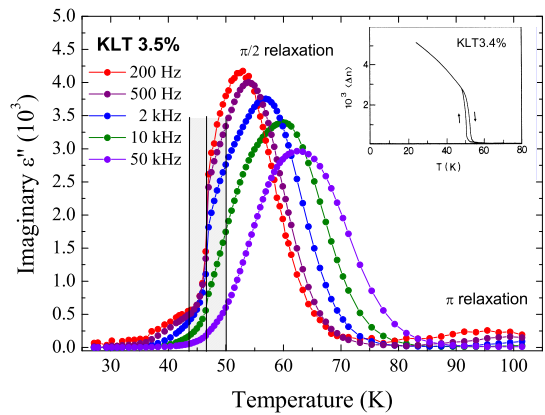}
\caption{Imaginary part of the dielectric constant of KLT3.5\% measured
at different frequencies (inset: birefringence of a different
KLT3.4\% crystal \cite{Azzini})}
\label{dielectric3.5}
\end{figure}

Fig.\,\ref{dielectric3.5} shows the imaginary part of the dielectric permittivity (absorption) of a K$_{\texttt 1-x}$Li$_{\texttt x}$TaO$_{\texttt 3}$ (KLT) crystal with a nominal concentration of 3.5\% Li, measured upon cooling as a function of temperature and at several frequencies. Two relaxation peaks  are visible. The small peak at $\sim95K$ corresponds to the 180$^\circ$ reorientation or $\pi$ relaxation and the large peak at lower temperature to the 90$^\circ$ reorientation or $\pi/2$ relaxation of the PNRs under the effect of the external ac field\,\cite{Pattnaik2}. The weaker strength of the $\pi$ relaxation in KLT3.5\% is due to the fact that the corresponding distortion (or elastic quadrupole) is the same for both crystallographic orientations (zero or $\pi$) of the PNRs. The transition to the tetragonal phase is evidenced by several experimental observations, probably the most direct of which are the sharp drop in the dielectric constant at T$_{c}$ $\approx47\,\mathrm{K}$ and a doubling of the phonon peaks detected in a neutron inelastic scattering study of KLT3.5\% \cite{Hennion}. The inset in Fig.\,\ref{dielectric3.5} also shows birefringence results obtained under continuous cooling and warming (no equilibration) in a KLT crystal with 3.4\% lithium\,\cite{Azzini}. Instead of a single curve, these reveal a narrow thermal hysteresis loop (see also \cite{Doussineau2}) which signals the existence of a two-phase region around the transition. Fluctuations between these two phases can therefore be expected in the vicinity of the transition (see later discussion on hetero-phase fluctuations)\,\cite{Gordon,Brookeman}. In this regard, it is useful to note that the width of the thermal hysteresis in the KLT3.4\% crystal is approximately the same as that of the hatched area on the  dielectric curve of the KLT3.5\% crystal, which represents the transition region ($T_c\pm4\,\mathrm{K}$) to be examined below. 

\begin{figure}[h]
\includegraphics[width=9cm]{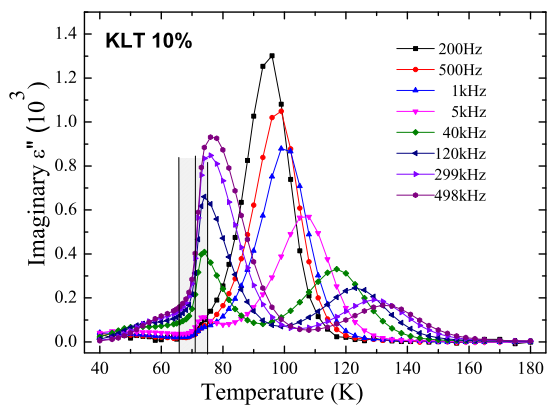}
\caption{Imaginary part of the dielectric constant of KLT10$\%$ measured
at different frequencies}
\label{dielectric10}
\end{figure}

For comparison, the corresponding dielectric results obtained on the KLT crystal with nominal x=10\% are presented in Fig.\,\ref{dielectric10}. For this higher concentration, the $\pi$ relaxation peak is much more prominent than in KLT3.5, possibly due to the larger size of the polar nanodomains and stronger strain fields. At the lower frequencies, the $\pi/2$ relaxation peak is barely visible, being mostly cut off by the intervening transition. Stated otherwise, the structural transition in KLT10\% intervenes at a higher temperature than that at which the $\pi/2$ relaxation peak would normally be observed if the transition did not occur.

\begin{figure*}
\includegraphics[angle=-90,width=16.5cm]{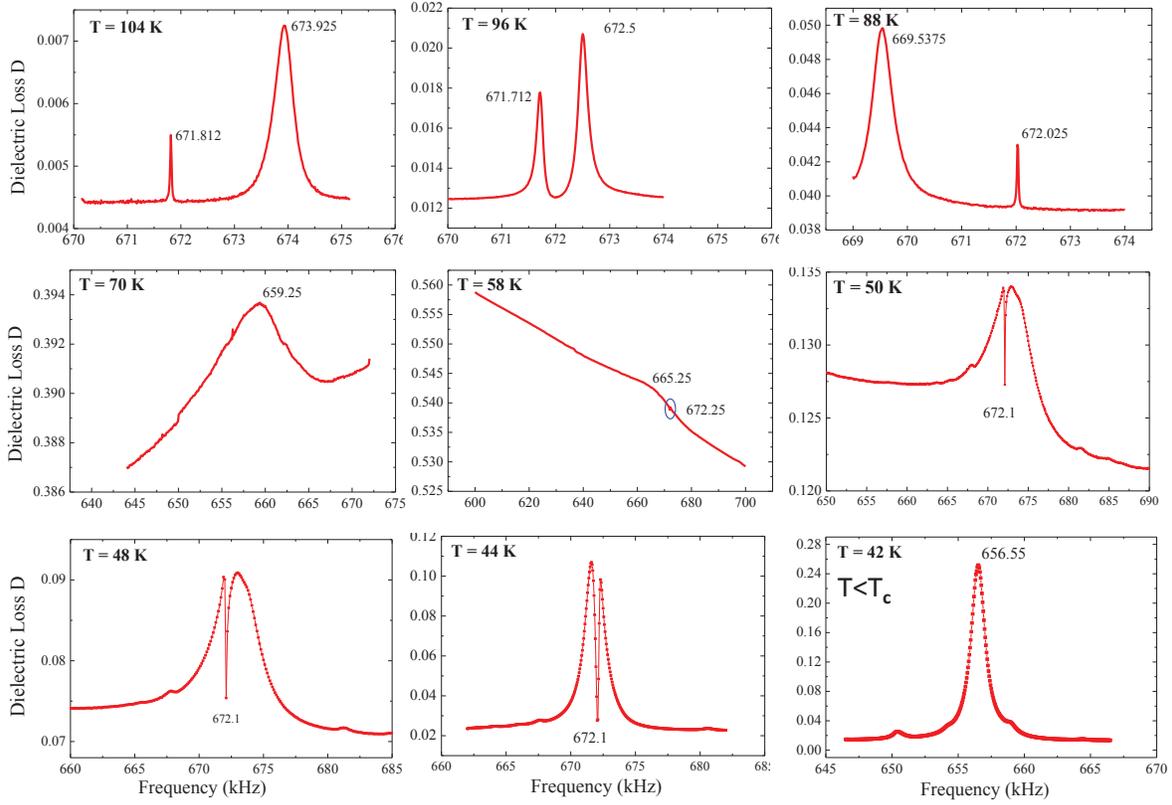}
\caption{Evolution with temperature of the dielectric loss, $D\equiv\epsilon''/\epsilon'$, of KLT3.5. At 70\,$\mathrm{K}$ the broad peak rides over the $\pi$ relaxation peak and is strongly damped at 58\,$\mathrm{K}$. At 50\,$\mathrm{K}$ both the broad and narrow resonance peaks have reappeared. And below the phase transition region, only the broad resonance is active (see explanation in text).}
\label{resonancesKLT3.5}
\end{figure*}

\begin{figure}[h!]
\includegraphics[width=9cm]{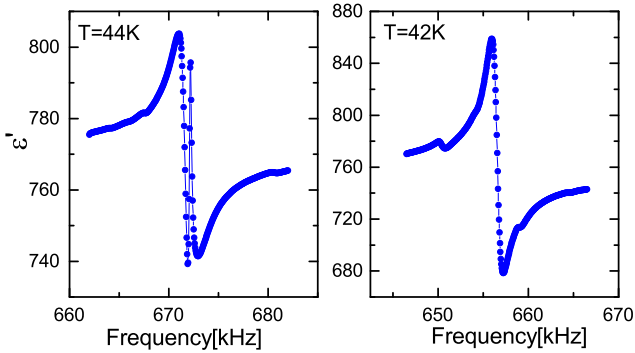}
\caption{Real part of the dielectric constant of KLT3.5 in the transition region (hatched area in Fig.\,\ref{dielectric3.5}) and just below}
\label{RealEpsilonKLT3.5}
\end{figure}

\begin{figure}[h!]
\includegraphics[width=8cm]{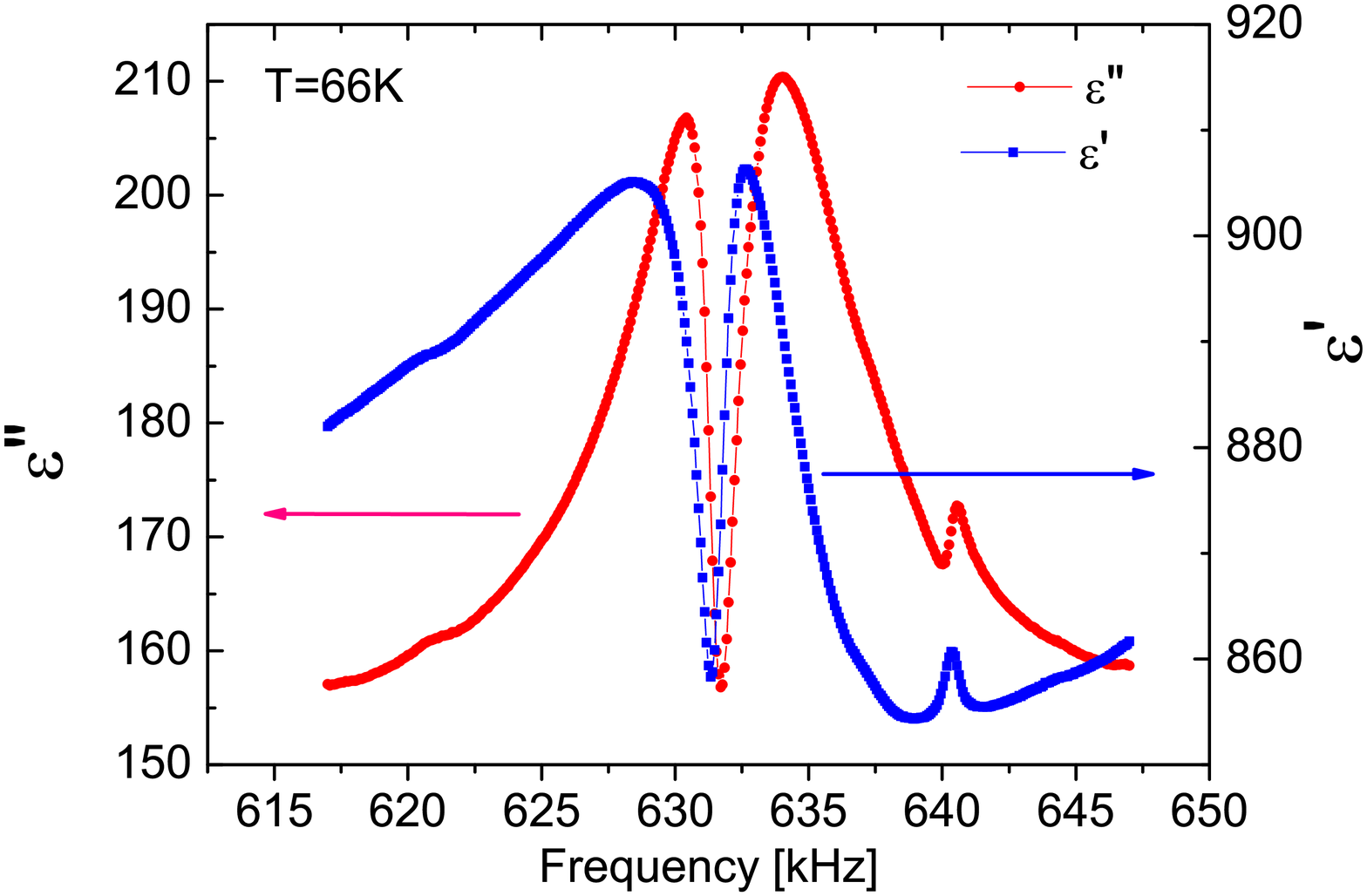}
\caption{Real and Imaginary parts of the dielectric constant of KLT10 in the transition region (hatched area in Fig.\,\ref{dielectric10})}
\label{resonanceKLT10}
\end{figure}

Relaxor ferroelectrics (RFs) also exhibit characteristic resonances, precisely in the same temperature region in which the PNRs are present and undergoing the $\pi/2$ and $\pi$ relaxations mentioned above (see also Ref.\,\cite{Pattnaik1}). For the measurements of these dielectric resonances, the same configuration was used as for the dielectric relaxation measurements save for a modest dc electric field ($\sim370\,\mathrm{V/cm}$) that was applied to partially align the PNRs and induce a small but non-zero macroscopic polarization. The frequency of the small measuring ac electric field was then swept through the resonance, the frequency of which can be calculated from the dimension of the crystal bar, $L$, the density, $\rho$, and the elastic compliance, $S$, as $\nu=\frac{1}{2L}{\sqrt{\frac{1}{\rho S}}}$. The results are presented in Fig.\,\ref{resonancesKLT3.5} in the form of the dielectric loss tangent, $D\equiv\epsilon''/\epsilon'$. Unexpectedly, not one but a pair of resonances is observed starting approximately at 120K, the temperature at which the (quasi-static or static) PNRs are known to appear as determined from independent measurements mentioned above (Raman \,\cite{Antonio} and diffuse neutron scattering \,\cite{Yong}). The more intense of the two resonances is broad and symmetric and the less intense is narrow and presents a characteristic asymmetry. It is important to note that the narrow resonance is not observed at room temperature. The frequencies of both resonances are found to fall within the same range as for length mode oscillations, associated with the longitudinal strain ($\epsilon_{11}$) of the bar, i.e. perpendicular to the direction of the applied electric field ($E_3$) and corresponding to the $d_{311}$ piezoelectric coefficient. The long dimension of the crystal bar samples was sufficiently different from the two others(width-thickness)  so as to exclude the possibility that the two resonances observed might correspond to two distinct modes of vibration. The thickness and width resonances were in fact observed separately at very different frequencies, by a dimensional factor of $8.6/5.5=1.6$ for the KLT3.5\% sample. The resonance frequencies on the KLT3.5 sample measured here were also found to be very close to those measured earlier in KTN samples when taking into account the respective sample dimensions.

\begin{figure}
\includegraphics[width=9cm]{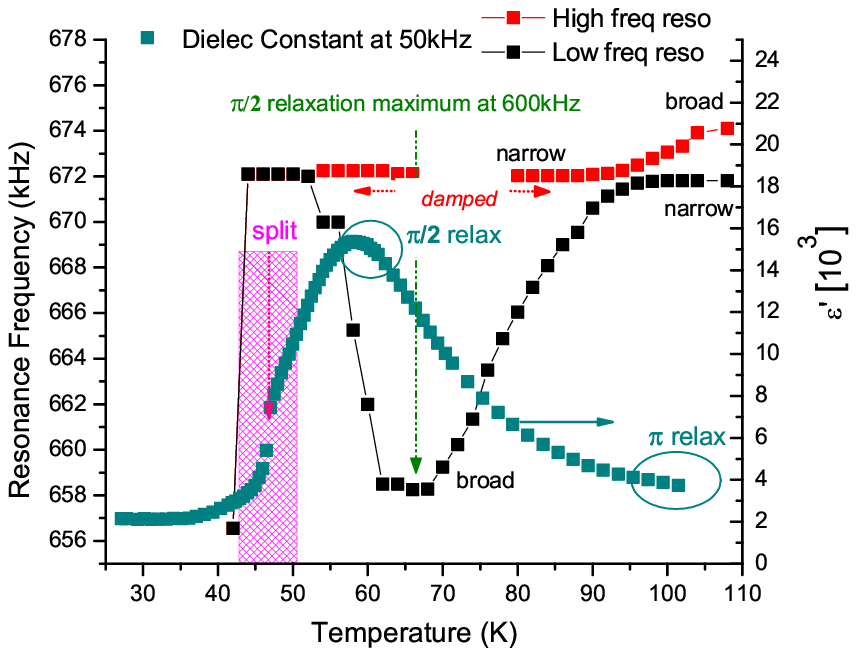}
\caption{Frequencies of the narrow and broad resonances and dielectric constant of KLT3.5 as a function of temperature. Note the anti-crossing of the broad and narrow resonance at T$\approx95K$. The dielectric constant is seen peaking in the temperature range of the minimum frequency of the broad resonance.}
\label{frequenciesKLT3.5}
\end{figure}

\begin{figure}
\includegraphics[width=9cm]{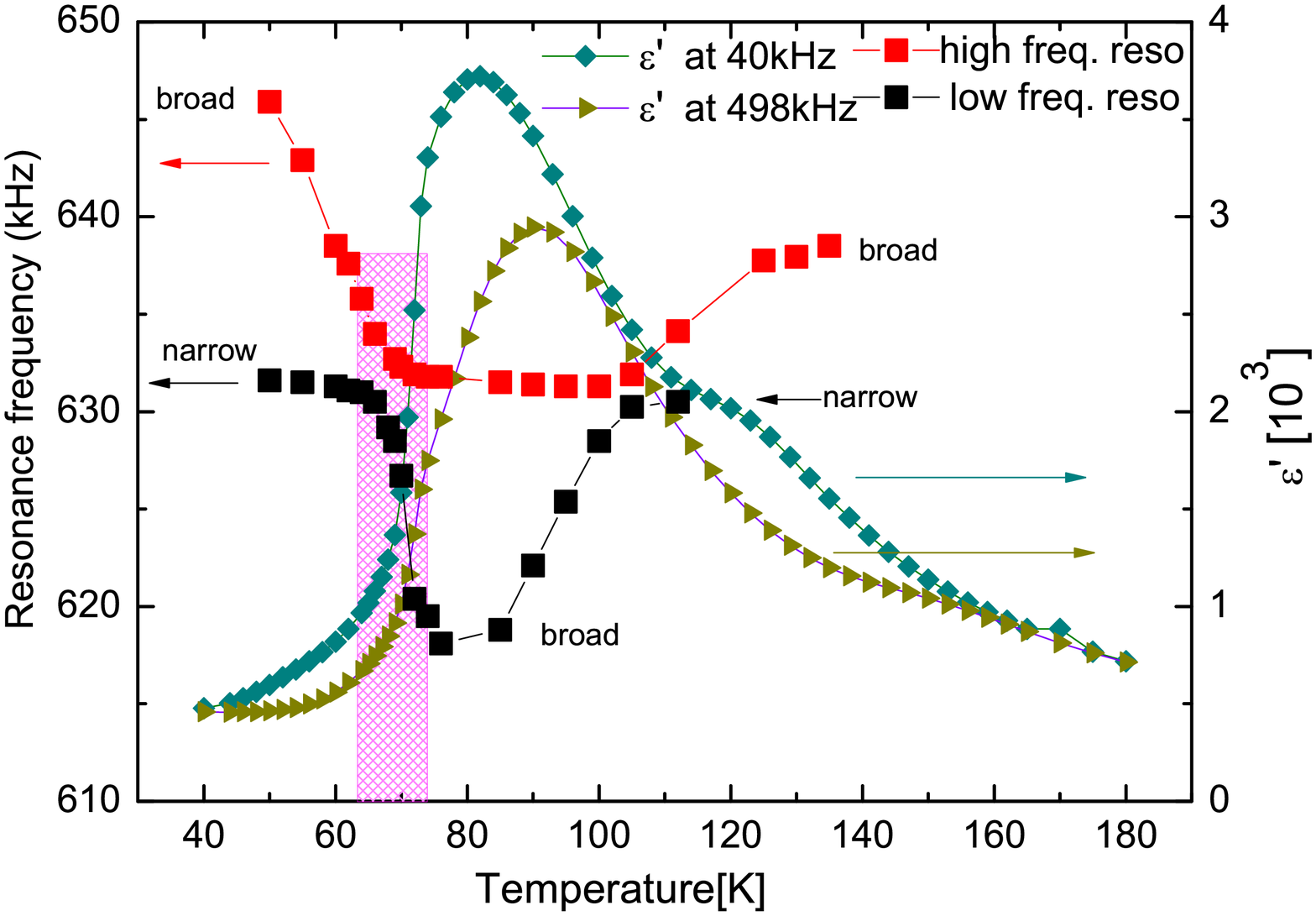}
\caption{Frequencies of the narrow and broad resonances and dielectric constant of KLT10 as a function of temperature. Note the anti-crossing of the broad and narrow resonance a T$\approx110K$. The dielectric constant is also seen peaking in the temperature range of the minimum frequency of the broad resonance.}
\label{frequenciesKLT10}
\end{figure}

Initially, the broad resonance is the much stronger one of the two. However, as it decreases in frequency and moves towards the narrow resonance, both its magnitude and width decrease, suggesting an energy exchange between the two resonances. At their point of closest approach (T$\approx{96K}$ in Fig.\,\ref{resonancesKLT3.5}), both resonance peaks are almost equal in strength. The corresponding real parts of the dielectric permittivity are also shown in Fig.\,\ref{RealEpsilonKLT3.5} at two temperatures within the transition range and below. Strikingly, below the transition a single resonance is observed. Very similar results are obtained in KLT10\% with two resonances above the transition but a single one below (not shown).
The frequency evolution of the broad and narrow resonances are shown in Figs.\,\ref{frequenciesKLT3.5} and \ref{frequenciesKLT10} for KLT3.5 and KLT10 respectively, together with the temperature evolution of the real part of the dielectric constant measured without an external dc field at 50\,$\mathrm{kHz}$ for KLT3.5 and at 40 kHz and 498\,$\mathrm{kHz}$ for KLT10. Also indicated in the same figures the position of the $\pi/2$ dielectric maximum can be seen to fall in the same frequency range as the resonances, suggesting an interaction between the two. Moreover, the frequency of the broad resonance reaches a minimum at approximately the same temperature as that of the maximum of the $\pi/2$ relaxation (when measured at the  same frequency as the resonance) and increases rapidly below. As will soon become apparent, the temperature positions of these relaxation maxima serve as important markers for the physical interpretation of the resonances. Accordingly, we identify three main temperature regions: region I of the $\pi$ relaxation peak, region II of the $\pi/2$ relaxation peak and region III around the transition. In region I, both resonances are present and grow in amplitude, with the narrow resonance becoming equal in strength to the broad resonance. One major difference however is that the frequency of the narrow resonance remains constant with temperature while that of the broad resonance decreases in frequency, (anti-)crossing over the narrow one at the temperature of the the $\pi$ relaxation maximum. In region II, the narrow resonance progressively disappears while the frequency of the  broad resonance continues to decrease with decreasing temperature, becoming strongly damped as it passes in the temperature range of the $\pi/2$ relaxation peak. In Region III, approaching the transition and with the $\pi/2$ relaxation now much too slow and therefore no longer active to dampen the resonances, both reappear, the broad resonance increasing again in frequency and decreasing in width. At 44$\mathrm{K}$, the two resonances are close to each other in frequency and overlap, resulting in a spectrum that superficially looks like a broad peak split in the middle. Finally, in the low temperature phase, at 42\,$\mathrm{K}$ for KLT3.5 and 68\,$\mathrm{K}$ for KLT10, only a single resonance is observed, which corresponds to the previously labeled broad resonance as we explained below. In the next Analysis section, we show that the resonance spectra presented above can be described equally well by two complementary models, each reflecting a different aspect of the resonance-relaxation dynamics. In the subsequent Discussion section, we then describe the physical mechanisms that explain the evolution of the dielectric resonance spectra.

\section{\label{sec:level3}Analysis}
In the present section, we show that the observed spectra in the three separate temperature regions identified above can be accurately described by either one of two complementary models, the first one purely classical and phenomenological and the second one semi-classical. Both models are shown to describe equally well the spectral shapes in regions I and III as an interplay of the resonances with a relaxation, the nature of this relaxation being different in the two regions. In the intermediate region II, the $\pi/2$ relaxation dominates and dampens the resonances. In what follows, we first establish the physical basis for the two models, then present each successively, and finally compare their predictions with the experimental results. Before proceeding however, several qualitative remarks can already be made in view of the results presented above to inform their interpretation: 1) the fact that both resonances appear when the PNRs are known to form\,\cite{Yong,Antonio} (breaking local inversion symmetry and inducing local piezoelectricity) at approximately 120\,$\mathrm{K}$ for KLT3.5 and 140\,$\mathrm{K}$ for KLT10, indicates that the PNRs must be the primary driver of these resonances, while the surrounding lattice remains cubic; 2) the observation of a pair of resonances rather than a single one in the frequency range for longitudinal oscillations of the bar samples suggests the existence of two distinct vibrational configurations for these oscillations, corresponding respectively to in-phase and out-of-phase oscillations of the PNRs and surrounding lattice or bar (see below); 3) the asymmetric shape of the narrow resonance peak does seem to suggest the existence of a coupling between these two modes of oscillation. In the following, two  models are proposed, one purely classical and the other semi-classical. Both are based on the above concept of two resonance modes coupled through a relaxation but they provide two complementary perspectives on the results. And both are shown to describe the experimental spectra very well, confirming the qualitative remarks above.\\

\textit{Purely Classical Model}\\

The observed resonance spectra can be described phenomenologically in terms of the dynamics of the well-known classical system of two damped oscillators coupled to each other, as expressed in Eq.\,\ref{Recoupling}\,\cite{Joeyong, Alzar}. In the present KLT case, one of the oscillators is the PNRs and the other the surrounding lattice. The even normal mode of this coupled system corresponds to the PNRs and surrounding lattice oscillating in phase (both simultaneously in extension or contraction) and the odd mode to them oscillating out-of-phase relative to each other.

\begin{align}
\begin{split}
\ddot{X}_1+\gamma_1\dot{X}_1+\omega_1^2X_1-\nu_{12}^2X_2&=a_1\exp({-i\omega t}) \\
\ddot{X}_2+\gamma_2\dot{X}_2+\omega_2^2X_2-\nu_{12}^2X_1&=0
\label{Recoupling}
\end{split}
\end{align}

These two normal modes can be coupled by an external force that flips the displacement (deformation) vector of one of the two oscillators and correspondingly the relative phase of its motion by $180^{\circ}$. Because the surrounding lattice or macroscopic bar sample is set into oscillations by the piezoelectric polar nano-domains, it is clear however that, initially, the ac field can only excite the even (primary) normal mode, with displacement $X_1$. The latter can then couple to the (odd) secondary mode with displacement $X_2$ through the polarization and strain reversal of the PNRs, and vice versa.
In the present KLT case at the higher temperature (Region I), the two modes are coupled through the $\pi$ relaxation or polarization reversal of the piezoelectric PNRs, accompanied by a change of sign of their strain state from expansion to contraction. And in the transition range (Region III), they are coupled by the relaxation of the surrounding lattice between its higher temperature cubic and its lower temperature tetragonal phase (heterophase relaxation), also accompanied by a change of sign of the strain from expansion to contraction and vice-versa. 

\begin{figure}[h!]
\includegraphics[width=7cm]{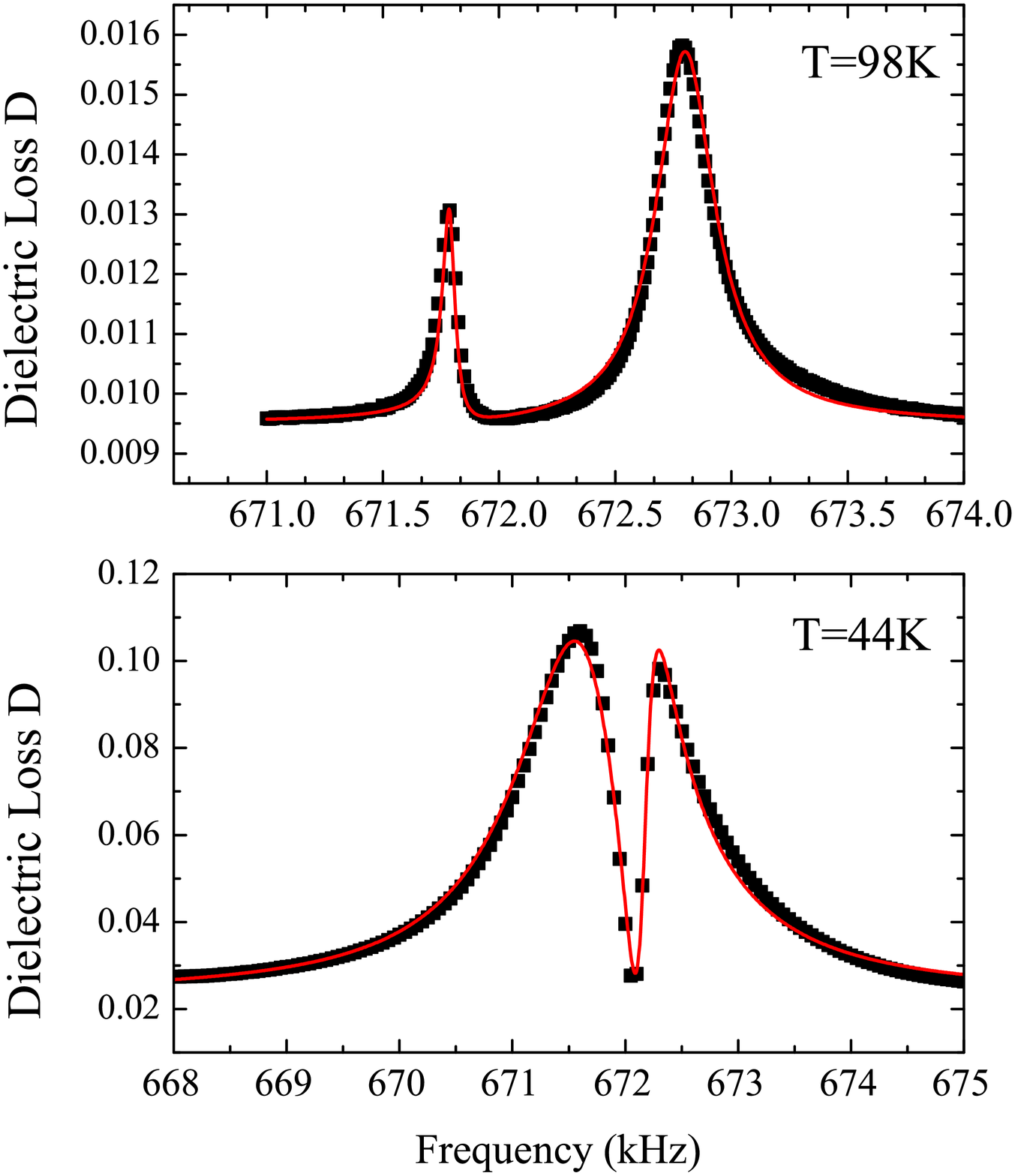}
\caption{Dielectric Loss spectra at 98K and 44K (same as in Fig.\,\ref{resonancesKLT3.5}) fitted with the two coupled oscillators model described by Eq.(\ref{Recoupling})}
\label{classicalresospectra}
\end{figure}

\begin{table}[h!]
\caption{Fit Parameters for the Purely Classical Model of two coupled damped harmonic oscillators}
\label{tab.1}
\begin{ruledtabular}
\begin{tabular}{lcr}
\textrm{Fit parameters} & \textrm{98K}&
\textrm{44K}\\ \hline \\
$a_1$ & $1.44 \times 10^6$ & $8.71 \times 10^7$ \\
$\omega_1/2\pi$(kHz) & 672.69 & 671.75 \\
$\omega_2/2\pi$(kHz) & 671.90 & 672.09 \\
$\gamma_1$ & 340.9 & 1587.2 \\
$\gamma_2$ & 32.1 & 17.8 \\
$\nu_{12}$ & $4.2 \times 10^8$  & $4.4 \times 10^8$  \\
background & 0.0095 & 0.0023  \\
\end{tabular}
\end{ruledtabular}
\label{simulparam}
\end{table}

The vibrational spectrum of such a well-known system does in fact reproduce very well the resonance spectra observed in both regions I and III. The experimental spectra and fits to the solution of Eq.\,(\ref{Recoupling}) at 98\,$\mathrm{K}$ and 44\,$\mathrm{K}$ are shown in Fig.\,\ref{classicalresospectra} and the fitting parameters are listed in Table\,\ref{simulparam}.
The quality of the fits to the experimental curves is excellent and the variation of the fitted values resulting from varying the starting values of the parameters is found to be less than $1\%$. As seen in Table\,\ref{simulparam}, the major differences between the two temperatures are \textit{i)}\, the much higher damping of the driven primary oscillator (broader resonance peak) and the slightly lower damping of the secondary oscillator at 44\,$\mathrm{K}$, and \textit{ii)}\, the smaller frequency separation and therefore greater overlap of the two resonances at 44\,$\mathrm{K}$ $\approx{350} kHz$ compared to $\approx{800} kHz$ at 98\,$\mathrm{K}$. Despite the coupling coefficient $\nu_{12}$ being almost the same at the two temperatures, the fact that the two resonance peaks overlap significantly at the lower temperature translates into a higher transition probability between the two modes. As explained below, such a higher transition probability at 44\,$\mathrm{K}$ can itself be explained by the proximity of the structural transition and the correspondingly much softer and deformable lattice.\\

\textit{Semi-Classical Model}\\

\begin{figure}[h!]
\includegraphics[width=6cm]{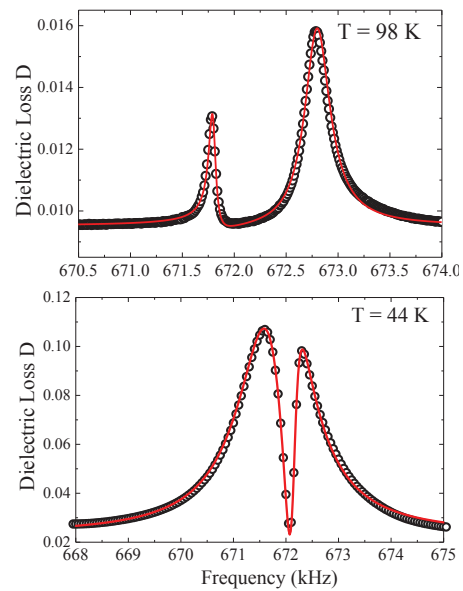}
\caption{Dielectric Loss spectra at 98K and 44K (same as in Fig.\,\ref{resonancesKLT3.5}), fitted with the ZR model in Eq.(\ref{loss})}
\label{semiclassicalresofit}
\end{figure}

As an alternative to the purely classical model above, the observed spectrum  can be equally well described by a semi-classical model which contributes a complementary physical perspective on the results. As described by Fano, an asymmetric lineshape, such as that of the narrow peak, from the coherent mixing of a vibrational transition from a ground state to a discrete excited state and a parallel transition to a continuum itself coupled to the excited state. The two parallel excitation paths thus result in an interference. The Fano resonance picture can be extended to the case of two separate excitations between discrete energy levels that are coupled to each other through a continuum. In the context of KLT, the two separate excitations are the even and odd modes and the continuum corresponds to the relaxation coupling the two. Such a situation was modeled several years ago by Zawadowski and Ruvalds\,(ZR)\,\cite{Zawadowski} for the case of two discrete and long wavelength optical phonons coupled to each other through pairs of acoustic phonons with wavevectors +$k$ and -$k$ and thus forming a continuum. The corresponding Green's function was taken to be purely imaginary, which is equivalent to a relaxation in the present KLT case. Given the one-to-one correspondence  between the vibrational configuration described by ZR and the present one, we can directly use the spectral function given in Eq.\,(12) of their paper to describe the dielectric loss spectra in KLT:

\begin{equation} \rho(\omega)=\frac{\epsilon''}{\epsilon'}=\frac{[Ag_{a}/2\Delta_{a}+Bg_{b}/2\Delta_{b}]^2}{1+
[g_{a}^2/2\Delta_{a}+g_{b}^2/2\Delta_{b}]^2}
\label{loss}
\end{equation}

in which $A$ and $B$ are the oscillator strengths of the two normal modes, $g_{\alpha}$ their respective coupling coefficients to the
relaxation and $\Delta_{\alpha}\equiv(1-\frac{\omega}{\omega_{\alpha}})$ the relative distance from their respective resonance frequencies. By contrast with the purely classical model described earlier by Eq.(\ref{Recoupling}), in the semi-classical ZR model both discrete vibrational normal modes are assumed to be driven by the ac field instead of just the (even primary) mode. The fitted spectra are presented in Fig.\,\ref{semiclassicalresofit} at 98\,$\mathrm{K}$, near the high temperature anti-crossing point in Region I, and at 44\,$\mathrm{K}$ close to the transition in region III.

\begin{table}[h!]
\caption{Fit parameters Semi-Classical Model of two discrete transitions coupled via a continuum}
\label{tab.2}
\begin{ruledtabular}
\begin{tabular}{lcr}
\textrm{Fit parameters}&
\textrm{98K}&
\textrm{44K}\\
\hline
A & 0.00169 & 0.0117 \\
B & 0.00106 & 0.0077 \\
$\omega_{a}/2\pi$(kHz) & 672.51 & 671.57 \\
$\omega_{b}/2\pi$(kHz) & 671.72 & 672.30 \\
$g_{a}$ & 0.01855 & 0.04033 \\
$g_{b}$ & 0.01451 & 0.028 \\
background & 0.0078 & 0.0231 \\
\end{tabular}
\end{ruledtabular}
\label{Fitvalues}
\end{table}

Here again, two seemingly very different spectra are fitted very well by the same model. The values of the fitting parameters are given in Table\,\ref{Fitvalues}. Contrary to the fitting results obtained with the purely classical model, the frequencies of the two resonances are found here to be practically the same at 98\,$\mathrm{K}$ and 44\,$\mathrm{K}$. In the present semi-classical model, the main difference between the spectra in Regions I and III is the larger values of both coupling coefficients or widths of the resonances, $g_a$ and $g_b$, which are twice as large at 44$\,\mathrm{K}$ as they are at 98\,$\mathrm{K}$, while remaining in the same ratio, 1.3-1.4. This doubling results in a extensive overlap of the two resonances. This extensive overlap of the two resonances, and the resulting much larger transition probability between the two modes, appears to be the essential common feature of the two models. In the purely classical model, this overlap stems from a significant increase in the damping/width of the even (primary) mode and the reduced frequency separation between the two modes while, in the semi-classical model, it originates from an equal increase in the coupling coefficients of both modes to the acoustic continuum or relaxation. It is also important to note that the coupling between the two oscillators is taken into account differently in the two models. In the purely classical model, the primary (driven) mode is coupled to the secondary (slave) mode through an implicit continuum (the $\pi$ or heterophase relaxation), with the coupling expressed in the damping coefficient. In the semi-classical model by contrast, both oscillators are coupled explicitly to a common continuum, each with its own coupling coefficient.\\

\section{\label{sec:level4}Physical Description and Discussion}
\begin{figure}[h!]
\includegraphics[angle=-90, width=9cm]{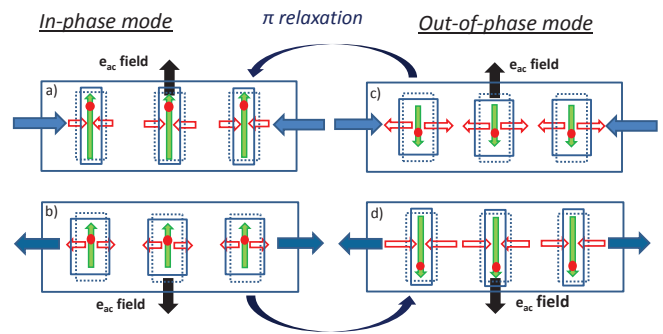}
\caption{Schematic representation of the modes of oscillations of the KLT crystal bar with PNRs partially aligned vertically in the dc electric field; the polarization of individual PNRs is represented by vertical green arrows and their deformation by horizontal open red arrows and dashed blue lines.
The red dot inside each PND represents the lithium ions whose cooperative $\pi$ relaxation is driven by the same ac electric
field that drives the resonances. The position of the red dots can be seen to be related to the nature of the deformation of the PND
(expansion or contraction). The deformation of the crystal bar is indicated by horizontal blue arrows. a),b) and c),d) correspond to the
two half-cycles of the in-phase and out-of-phase oscillations respectively.}
\label{KLTmodelpi}
\end{figure}

\begin{figure}[h!]
\includegraphics[width=7cm]{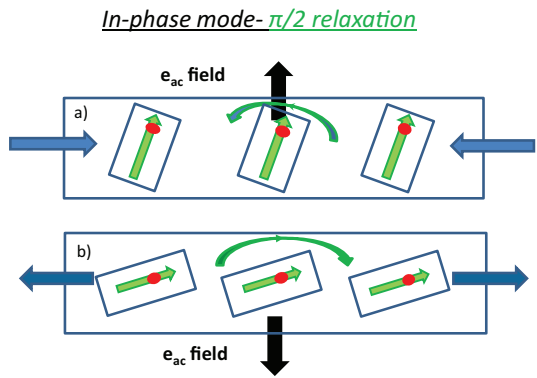}
\caption{The two half-cycles of PNRs undergoing a $\pi/2$ relaxation, which broadens the broad resonance and suppresses the narrow resonance; the features in this figure are the same as in the previous figure}
\label{KLTmodelpi2}
\end{figure}

\begin{figure}[h!]
\includegraphics[angle=-90, width=10cm]{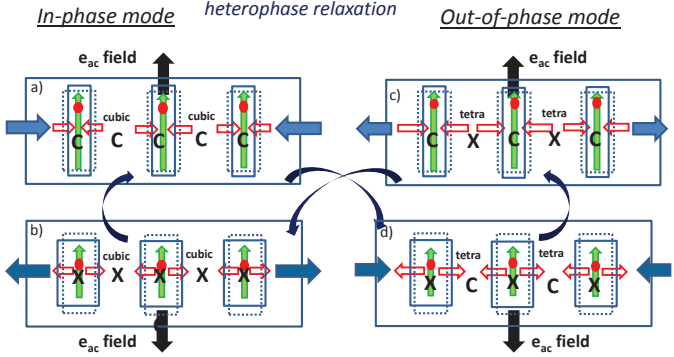}
\caption{Normal modes of oscillations of the PNRs and surrounding lattice in the transition region. The color codes are the same as in Region I representation. At the lower temperature of the transition however, both the $\pi$ and $\pi/2$ relaxations are inactive and the PNR polarization remains aligned along the dc field. C and X designate contraction and expansion respectively and the crystal symmetry of the surrounding lattice is also indicated. Following the ac field, the system evolves along the path a)-d)-c)-b)-a). In the left in-phase mode (b-a), the surrounding lattice between PNRs is cubic. But its contraction in a) induces a transition to the tetragonal phase in d). Similarly, its expansion in c) induces a transition back to the cubic phase in b).}
\label{KLTmodelHetero}
\end{figure}

\indent The physics underlying the models presented above can be tentatively illustrated as in Figs.\,\ref{KLTmodelpi} (Region I), \ref{KLTmodelpi2} (Region II) and \ref{KLTmodelHetero} (Region III). Although these pictures are an attempt at describing the physical phenomena observed and successfully modeled in the previous section, their validity rests on the coherence of the description across the three regions. In each figure, the PNRs are assumed to be aligned, at least partially, by the dc field while the ac field excites the resonances and relaxation. In region I, at high temperature, the even (in-phase) and odd (out-of-phase) normal modes of the PNRs-surrounding lattice system are coupled via the $\pi$ relaxation of the lithium ions (red dots in Fig.\,\ref{KLTmodelpi}) that switches the polarization of the PNRs (green arrows) by $180^{\circ}$ and, correspondingly, their  piezoelectric deformation from expansion or dilation to contraction. It should however be obvious that, initially, the ac field can only excite the coupled system in its even or in-phase mode since it is the piezoelectric deformation of the PNRs that initially drives the surrounding lattice and bar into oscillations. Only once the system has been set oscillating in the in-phase mode of vibration can it transition back and forth between the two modes through a reversal of the PND polarization and associated piezoelectric deformation triggered by the $\pi$ relaxation.

One important aspect of the observed dynamics in Region I is that it is coherent, since the same ac field that excites the piezoelectric resonance of the PNRs also triggers their $\pi$ relaxation and accompanying $180^{\circ}$ polarization reversal. The $\pi$ relaxation itself is therefore assisted by the piezoelectric deformation of the PNRs, which reduces the potential barrier for the $\pi$ relaxation of the lithium ions and reversal of the PNRs polarization. And this process is clearly more effective for the in-phase or even mode since the deformations of the PNRs and surrounding lattice are then both of the same sign (see evolution from b) to c) or d) in Fig.\,\ref{KLTmodelpi}). A stronger coupling to the relaxation means a higher damping of the in-phase mode and a broader resonance peak, as indeed observed.

\indent Region II is the temperature region  within which the $\pi/2$ relaxation reaches its maximum amplitude, $\omega\tau_{\pi/2}=1$ (see Fig.\,\ref{dielectric3.5}). Unlike in Region I, the $90^{\circ}$ reorientation of the PNRs in region II does not couple the two oscillation modes to each other, and the out-of-phase mode (narrow asymmetric resonance) therefore vanishes. Additionally, and as seen in Figs.\,\ref{resonancesKLT3.5},  \ref{frequenciesKLT3.5} and \ref{frequenciesKLT10}, the $\pi/2$ relaxation crosses over the frequency of the broad resonance, strongly damping it and depressing its frequency (as for a damped harmonic oscillator with increasing damping). The effect of the $\pi/2$ relaxation is illustrated in Fig.\,\ref{KLTmodelpi2}. This mechanism explains both the increased damping and rapid frequency decrease of the broad resonance and the disappearance of the narrow resonance, which only exists through its coupling to the broad resonance via the $\pi$ relaxation. 
At lower temperatures, the $\pi/2$ relaxation itself slows down and in turn becomes inactive. As a result, the broad resonance is no longer damped, it recovers an even larger amplitude than before and its frequency increases again. The model used here to describe the evolution of the resonances in Region II is therefore fully consistent with the model used in Region I, itself based on the interaction of the resonance with a relaxation, $\pi$ in Regions I and $\pi/2$ in Region II.\\

\indent In region III, although the $\pi$ and $\pi/2$ relaxations are no longer active, the two resonances nevertheless reappear, now strongly overlapping. The fact that the same two models are able to reproduce the experimental spectra in Region I and III indicates that a similar generic explanation must apply in both, in terms of two discrete oscillators (even and odd normal modes) coupled through a continuum (a relaxation). However, as already hinted above, the physical nature of the relaxation coupling the two modes is not the same in both regions. In Region I, the even and odd modes correspond to the in-phase and out-of-phase oscillations of the PNRs-surrounding lattice system, coupled to each other through the $\pi$ relaxation of the PNRs. But Region III lies well below the peak temperature of both relaxations, which are therefore inactive. The two modes must now be coupled through a different kind of relaxation, the nature of which is revealed by two observations: \textit{i)} Region III straddles the structural transition, as shown by the hatched area in Fig.\ref{dielectric3.5}, and \textit{ii)} the thermal hysteresis seen in the inset of the same figure shows that Region III is a region in  which the high and low temperature phases are metastable on some time scale, or relax slowly from one to the other. These two observations suggest that the relaxation that is active in Region III is that of the surrounding lattice between its high temperature (cubic) and low temperature (tetragonal) phases, otherwise called "heterophase relaxation".

Heterophase fluctuations are indeed observed near weakly first order transitions, where they are due to the presence of precursors of a low temperature phase within a high temperature equilibrium phase and vice versa.\,\cite {Cook,Gordon} And they are quite naturally expected to occur in relaxors since the PNRs do indeed represent stable precursors of the low temperature phase, already present above the transition. Because the PNRs are intrinsically piezoelectric, modulation of their polarization by the ac field leads to a modulated stress on the surrounding lattice which, being already near a structural instability in the vicinity of the transition, can easily be made to transform from the cubic to the tetragonal phase and vice-versa. This transformation is necessarily accompanied by a phase change in the oscillations of the surrounding lattice relative to the PNRs, or coupling of the even/in-phase mode and the odd/out-of-phase mode as in Region I. Here however, instead of the lithium ions or PNR polarization relaxing, it is now the lattice that is relaxing between the cubic and tetragonal phases. Moreover, because the heterophase relaxation is induced by the piezoelectric oscillations of the PNRs, it is also coherent with the latter. Hence, the two coupled-mode picture that explains the dynamics in Region I is also valid in Region III, although the physical nature of the relaxation coupling the two modes is different in the two Regions. Fig.\,\ref{KLTmodelHetero} represents an attempt to illustrate the likely sequence  for the resonance-relaxation process in which the ac field again modulates the polarization of the PNRs whose deformation drives the heterophase relaxation of the surrounding lattice and ultimately the macroscopic bar oscillations.

Starting with half-cycle $\textit{a)}$ and the cubic lattice, the combination of the dc and ac fields enhances the polarization of the PNRs which contract (C), resulting in a contraction of the surrounding lattice and bar sample. The latter contractions then induce the cubic-to-tetragonal transformation of the lattice in half-cycle $\textit{d)}$, which maintains the bar in contraction while the ac field has now reversed and the PNRs are expanding. The PNRs-surrounding lattice system now oscillates in the odd mode. In half-cycle $\textit{c)}$, with the reversal of the ac field but still in the out-of-phase mode, the PNRs now contract while the surrounding lattice expands, inducing the reverse tetragonal-to-cubic transformation of the lattice and coupling the odd mode back to the even mode in half-cycle $\textit{b)}$. Here we note that the two half-cycles $\textit{a)}$ and $\textit{b)}$ of the even mode in the cubic lattice are identical to those in Fig.\ref{KLTmodelpi}, but that the two half-cycles $\textit{c)}$ and $\textit{d)}$ of the odd mode differ from the previous ones due to the cubic-tetragonal phase transition taking place as a result of the heterophase relaxation. The proposed model is again consistent with the lower frequency and much higher damping coefficient of the even (in-phase) mode.\\

\indent We now address the question of the coherence between the in-phase and out-of-phase oscillations (even and odd modes) of the system, a coherence that is essential to the efficient transfer of energy between them, and the question of why this coherence is not destroyed by the $\pi/2$ relaxation in Regions I and III. In the higher temperature Region I, the $\pi/2$ relaxation is very rapid ($\omega\tau_{\pi/2}\ll{1}$), the time-averaged fraction of PNRs aligned along the dc field axis is therefore statistically constant and the relaxational dynamics is dominated by the $\pi$ relaxation.  In Region III, the $\pi$ and $\pi/2$ relaxations are no longer active on the time (frequency) scale of the measurement and the fraction of PNRs aligned along the dc field is again constant, now in an absolute rather than statistical sense. As to the nature of the two coupled oscillators and the physical coupling mechanism, it is different in Regions I and III. As explained earlier, the two modes in Region I are the in-phase and the out-of-phase oscillations of the PNRs and surrounding lattice or bar system, coupled to each other via the $180^{\circ}$ or $\pi$ relaxation. In region  III, where the $\pi$ relaxation is no longer active, the relaxation that couples the two oscillation modes is now the structural transformation of the lattice driven by the piezoelectric PNRs. It is important to emphasize that the channeling of the lithium ions and polarization reversal in Region I on the one hand, and the structural transformation of the surrounding lattice in Region III on the other, are both assisted by the piezoelectric deformation of the PNRs, thus ensuring coherency between the in-phase and out-of phase oscillations. In region III, this deformation is magnified by the heterophase relaxation, as suggested by a 4.6 fold increase in the damping of the driven primary oscillator in the purely classical model and the doubling of both coupling coefficients in the semi-classical model.\\

\indent The necessary condition for the observation of coupled resonances such as those reported above in KLT are the presence of piezoelectric polar nanodomains with orientational degrees of freedom. These are in fact the characteristic features of relaxor ferroelectrics. It is therefore not surprising that similar resonances have been observed in other relaxors as well, KTa$_{\tt 1-x}$Nb$_{\tt x}$O$_{\tt 3}$ (KTN), PbMg$_{\tt 1/3}$Nb$_{\tt 2/3}$O$_{3}$ (PMN) \cite{Pattnaik1} and PbZn$_{\tt 1/2}$Nb$_{\tt 1/2}$O$_{\tt 3}$ (PZN).
The present report on the PNR-related resonances observed in KLT, and their analysis and interpretation, should therefore contribute broadly to a better understanding of the multiscale dynamics in relaxor ferroelectrics, explaining how their macroscopic properties emerge from their structural and dynamical properties at the nano level.\\

\begin{figure}[h!]
\includegraphics[angle=-90, width=11.5cm]{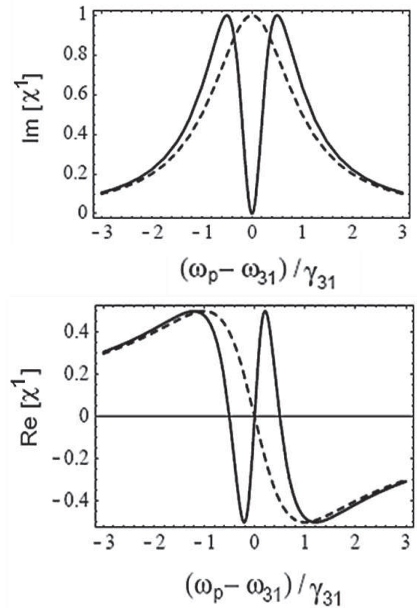}
\centering
\caption{Imaginary (top) and Real (bottom) parts of the optical susceptibility of an atomic rubidium vapor \cite{MF}. $\omega_p$ is the probe (laser) frequency,  $\omega_{31}$ and $\gamma_{31}$ the resonant frequency and damping respectively of the primary transition (in KLT, (1) would be the polarized state and (3) the in-phase strained state of the PNRs-surrounding lattice system with state (2) corresponding to the out-of-phase strained state). The dash curve corresponds to the susceptibility of a usual two state system (1-3), which does not exhibit EIT.}
\label{EIT}
\end{figure}

\indent Besides their contribution to a better understanding of relaxor ferroelectrics, the above results may also be of a general interest in Condensed Matter Physics. Relaxor ferroelectrics are but one example of what can be called coherent nanocomposites. Such systems are
characterized by a nanometer scale local order that is structurally coherent with the surrounding lattice, as in relaxor ferroelectrics. Similar types of phenomena as those described in the present paper are likely to be observed for instance in nanocomposite magnetic systems.\cite{Kimura} The resonance phenomena reported here may also be of interest at a more general physical level. They are indeed conceptually similar to phenomena observed in very different fields of physics, and in particular electromagnetically induced transparency (EIT) in atomic physics. The dielectric susceptibility spectra reported above in the relaxor KLT near the phase transition are indeed almost identical to the optical susceptibility spectra resulting from EIT in atomic vapors and reproduced here in Fig.\,\ref{EIT} for rubidium from Ref.\cite{MF} (compare with the KLT spectra in Figs.\,\ref{resonancesKLT3.5} and \ref{RealEpsilonKLT3.5}. The physical model used to describe the resonance phenomenon in KLT can in fact also be described semi-classically by analogy with the formalism of EIT for an atom with three discrete states (a ground state (1) and two coupled excited states (2,3)) exhibiting two closely spaced lifetime-broadened resonances that decay to the same continuum \cite{Harris} (see also \cite{Marangos}). In KLT, the ground state corresponds to the polarized state and the two excited states to the in-phase and out-of-phase strained states of the PNR-surrounding lattice system. The energy width of the two excited states is associated with the damping of the oscillators in the classical model and with the coupling strength of the two oscillators to the continuum in the semi-classical model, both leading to greater overlap and increased coherence between the two resonances. The correspondence of the dynamics of KLT and other relaxors with EIT will be further explored in a subsequent paper.\\

In conclusion, we have reported the observation of pairs or coupled resonances in the relaxor ferroelectric
K$_{\tt{1-x}}$Li$_{\tt x}$TaO$_{\tt 3}$ (KLT) as well as in others. These resonances provide a window into the multiscale dynamics
of complex oxides, from the nano- to the macro- scale. They are shown to be associated with two distinct oscillating configurations
or normal modes of the nanocomposite PNRs-surrounding lattice system, coherently coupled to each other via a relaxation. The observed spectra exhibit characteristic Fano lineshapes which evolve rapidly through three temperature ranges, due to the
complex interactions between resonances and relaxations. Despite this rapid evolution, the resonance spectra are explained equally well
and fitted over the entire temperature range using either one of two models, a purely classical or a semi-classical model, each highlighting a particular aspect of the system. Similar spectra are observed in other relaxors, pointing to the generality of these results. Finally, it is worth mentioning that other types of measurements have also revealed Fano lineshapes in disordered relaxor ferroelectrics, such as the optic $TO_2$ phonon in the Raman spectrum of Li-doped $KTa_{\tt{1-x}}$Nb$_{\tt x}$O$_{\tt 3}$ (KLTN) \cite{Kojima} or the optical phonon mode of the Zr sublattice in the imaginary dielectric spectrum of $BaZr_{0.5}$Ti$_{0.5}$O$_{\tt 3}$ (BZT).\cite{Wang}

\begin{acknowledgments}
The early part of this work was supported by a grant from the US Department of Energy, Office of Basic Energy Sciences, DE-FG02-06ER46318
and the NSF-REU program (JL). We also wish to thank Dr. L.A. Boatner from the Oak Ridge National Laboratory for providing the crystals used in the present study. Special thanks to J.F. Scott for pointing out the ZR paper to us.
\end{acknowledgments}

\end{document}